%% file: paper.tex
\documentclass[a4paper]{amsart}

\usepackage{prelim2e}
\usepackage{sty/fixme}

\usepackage{ucs}
\usepackage[utf8x]{inputenc}
\usepackage[T1]{fontenc}
\usepackage[english]{babel}
\usepackage[hyphens]{url}%
\usepackage{ifpdf}

\usepackage[printonlyused]{acronym}
\usepackage{authblk}
\usepackage{booktabs}
\usepackage{color}
\usepackage[shortlabels,inline]{enumitem}
\usepackage{float}%
  \floatstyle{ruled}%
  \newfloat{Algorithm}{tp}{lsa}%
\usepackage[final]{graphicx}
\usepackage{sty/listings}
\usepackage{sty/mathitbf}
\usepackage{rotating}
\usepackage{xcolor}

\usepackage{hyperref}
  \hypersetup{%
    colorlinks,
    linkcolor=black,
    citecolor=blue,
  }%
  \ifpdf\hypersetup{%
    pdftitle={Batch-oriented software appliances},%
    pdfauthor={R. Murri and S. Maffioletti},%
    pdfkeywords={Virtualization, User-Mode Linux, Software Appliance, Grid Computing},%
    pdfstartpage=1,%
  }\fi%

\begin{document}
\renewcommand{\paragraph}[1]{{\em #1.}}

\title{Batch-oriented software appliances}
\author{Riccardo Murri}
\author{Sergio Maffioletti}
\affil{
  \email{riccardo.murri@gmail.com}, \email{sergio.maffioletti@gc3.uzh.ch}\\
  Grid Computing Competence Center,\\
  Organisch-Chemisches Institut,\\
  Universität Zürich,\\
  Winterthurerstrasse 190, CH-8006 Zürich,\\
  Switzerland.\\
}%
\date{Mar.~6, 2012}
\maketitle

\begin{abstract}
  This paper presents AppPot, a system for creating 
  Linux software appliances. AppPot can be run as a
  regular batch or grid job and executed in user space, and
  requires no special virtualization support in the infrastructure.

  The main design goal of AppPot is to bring the benefits of a
  virtualization-based \acs{IaaS} cloud to existing batch-oriented
  computing infrastructures.  

  In particular, AppPot addresses the application deployment and
  configuration on large heterogeneous computing infrastructures:
  users are enabled to prepare their own customized virtual appliance
  for providing a safe execution environment for their
  applications. These appliances can then be executed on virtually any
  computing infrastructure being in a private or public cloud as well
  as any batch-controlled computing clusters the user may have access
  to.

  We give an overview of AppPot and its features, the technology that
  makes it possible, and report on experiences running it in
  production use within the Swiss National Grid infrastructure
  \acs{SMSCG}.

\end{abstract}


\section{Introduction}
\label{sec:intro}

Application deployment and configuration on large heterogeneous
systems is a complex infrastructure issue that requires
coordination among various system administrators, end users as well as
operation teams. This is further complicated when it comes to scientic
applications that are, most of the time, not supported on many Linux
distributios.

Virtualized infrastructures and software appliances provide an
effective solution to these problems but do require a specific
infrastructure and a usage model that is markedly different form the
batch-oriented processing that is still the backbone of scientific
computing.

This paper presents a system (nicknamed ``AppPot'') to bring the
benefits of virtualization-based \acs{IaaS} clouds to existing
batch-oriented computing infrastructures. 

AppPot comes in the form of a set of \acs{POSIX} shell 
scripts that are installed into a GNU/Linux system image, and modify
its start-up sequence to allow controlling task execution via the
kernel boot command-line.  Another utility is provided to invoke an
AppPot system image from the shell command line, and request it to
execute a given command.  The software is freely available from
\url{http://code.google.com/p/apppot}.

This combination effectively turns AppPot into a technology for
constructing
\href{http://en.wikipedia.org/wiki/Software_appliance}{\emph{software
    appliances}} that can also run as batch jobs in a local cluster or
grid computing system.  Pairing this with the \acl{UML} virtualization
system \cite{uml,uml:website}, AppPot appliances require (almost) no
support from grid and cluster systems administrators.  This
effectively allows use cases that have so far made \acs{IaaS} cloud
infrastructures attractive for end-users.

The rest of the paper is organized as follow:
Section~\ref{sec:motivation} explains the motivation behind the
project; Section~\ref{sec:usecases} presents typical usecases and the main
restrictions that AppPot helps addressing. Section~\ref{sec:uml} recaps
the main features of \ac{UML}, the virtualization technology that
allows AppPot to run without ``root'' user privileges.
Section~\ref{sec:architecture} presents the architectural
details as well as the functional specifications. 
Section~\ref{sec:usage} elaborates on how the presented
use cases were implemented using AppPot, and reports on the observed limitations.

\section{Motivations}
\label{sec:motivation}

A problem that has traditionally plagued batch computing
infrastructures is the deployment of software applications: in a
centrally-administered system, all requests for software installation
must be acted upon by the systems administrator.  This is less of an
issue in local computing clusters, where users can usually freely
install software in their own home directories, but scales up to a
significant administration and communication problem in a large
computational grid.

In addition, some software packages (notably, many scientific codes)
require complex installation procedures or provide scarce
documentation, so that specific expertise is needed to properly
install and configure the application.  In a grid infrastructure, this
poses an organizational scalability problem again: all systems
administrators must be conversant with the installation procedures of
every software piece.  Indeed, this issue has been tackled by many
grid infrastructures, either by providing access only to a fixed
restricted set of software applications, or by requiring clearance
procedures to elevate a subset of the users (``\acs{VO} operators'')
to the privilege level needed to install software on the execution
nodes.  The first solution restricts the possibilities of users to
exploit the infrastructure to its full potential; the second one
burdens end-users with additional tasks.

While the installation of publicly-available software packages is
generally negotiable in some way, central administration of software
becomes overly impractical when users need to deploy an application
they are developing themselves.  In this case, the code changes very
frequently, and it is just not feasible to issue an installation
request for every revision.  Still, users might need to execute
validation tests and regression suites, which in the case of scientific
applications can take hours to run; this is indeed a perfect use case
for batch jobs.

Leveraging the \acl{UML} virtualization system, AppPot
appliances can run on grid and local clusters as regular batch system
jobs, without the need for sysadmin support or root access.
This solves both the aforementioned problems:
\begin{itemize}
\item AppPot software appliances are a way to implement generic
  application deployment on a computational grid, and especially to
  enable users to provide their own software to the computing cluster:
  a complete AppPot appliance consists of three files, that can be
  copied to the execution cluster with any available mechanism,
  including the ``stage in'' facilities provided by most grid and
  cluster batch systems.
\item Users can use an AppPot \ac{VM} on their computer for coding,
  and then run the same \ac{VM} as a Grid jobs or in a Cloud \ac{IaaS}
  infrastructure for larger tests and production runs.
\end{itemize}





\section{Goals and use cases}
\label{sec:usecases}

The following scenarios are meant as an illustration of AppPot's
intended use cases.  Throughout the paper, we shall describe how the
requirements from these use cases translate into design decisions for
AppPot, and how well the goals have been met.

\subsection{Deployment of complex applications}
\label{sec:usecase-deployment}

Some software packages (notably, many scientific codes) require
complex installation procedures.  Typical problems may be roughly
summarized in the following categories:
\begin{enumerate}[\itshape (R1)]
\item\label{R1} 
  The application depends on software that is not readily
  available on the host operating system, or 
  the application depends on software or a specific configuration
  that conflicts with other systems settings.
\item\label{R2} 
  The application has a complex or non-standard compilation
  procedure, and the documentation is scarce.  For example, the
  installation procedure may require that some files are hand-edited
  and placed into specific locations, but the documentation is not
  clear about the content and format of these files, or on how to
  adapt the examples to new systems.
\end{enumerate}
In a grid infrastructure, this poses an additional problem of scale:
all systems administrators must be conversant with the installation
procedures of every software piece, and every application must be
compatible with all the computing systems available in the
infrastructure. 

It is well-known how virtual appliances allow solving these problems; 
however, virtualization systems that require ``root'' privileges for
operation pose a security problem, since the attack surface of an
appliance is larger than the one of a unprivileged process.  This
translates into the following requirement:
\begin{enumerate}[\itshape (R1),resume]
\item\label{R3} Ensure that deployed appliances cannot do harm to
  the system.\footnote{%
    If users have shell access to a system, as is commonly the case
    with cluster installations, then they can already run arbitrary
    code, so the requirement practically weakens to ensuring that
    appliances cannot do \emph{more} harm than a regular user process
    already can.
  } (See, e.g., \cite{ormandy:2007})
\end{enumerate}

\subsection{Running self-developed code}
\label{sec:usecase-development}

A large fraction of research groups are developing their own software
applications;%
oftentimes for computational experiments that are ephemeral, or limited
in scope to a local group or niche community.
  
All the problems illustrated in the previous use case still
apply, and a few additional features have to be considered here.
Namely, small updates to the software appliance are frequent and
further progress in the code may depend on the outcome of the tests
run on the updated appliance, so:
\begin{enumerate}[\itshape (R1),resume]
\item\label{R6} 
  It should be easy for users to prepare appliances to
  run different versions and branches of the same software, in order
  to experiment with algorithmic variants.
\item\label{R7} 
  It should be easy for users to add, remove and change software
  dependencies from the appliance, as the code grows and evolves.
\item\label{R4} 
  The deployment procedure should not involve any one else than
  the code author: it should be completely controlled and initiated by
  the user.
\item\label{R5}
  The deployment procedure should be as fast as possible not to
  interfere with the software write/test cycle.
\end{enumerate}

\section{User-Mode Linux}
\label{sec:uml}

The fundamental ingredient for AppPot is the \acl{UML} virtualization
system \cite{uml,uml:website}.  We shall therefore recap here the main
points of its architecture and the features that make it suitable for
AppPot.

\acf{UML} consists of a modified Linux kernel (guest) that runs as a
userspace process within another Linux system (host); being a regular
Linux kernel in almost every other aspect, \ac{UML} can run any Linux
distribution with any configuration.  The main difference of \ac{UML}
relative to other (para)virtualization solutions is that \acl{UML} can
only run a Linux guest inside Linux host: no other \ac{OS} is
supported.

\ac{UML} supports many of the features that make \acs{OS}
virtualization products attractive for building appliances.
AppPot leverages the following ones:
\begin{itemize}
\item Any file in the host system can be mapped to a block device in
  the guest system.  \ac{UML} has a copy-on-write feature, wherein all
  writes to the filesystem are written to a separate file, so that a
  single filesystem image can be shared by many concurrent \ac{UML}
  instances.
\item Portions of the host filesystem can be grafted in the guest
  filesystem, with full read/write access.\footnote{%
    Normal access control on the host filesystem applies: for
    instance, a guest run by an unprivileged user cannot modify files
    that are owned by the super-user ``root''.  }
\item The additional helper program \texttt{slirp} (see
  \cite{slirp:wikipedia,knoble:1996:slirp}) enables use of the host's
  \acs{TCP}/\acs{UDP} networking from within the guest system\footnote{%
    \ac{UML} is actually extremely flexible in terms of network device
    support, which has made it very popular as a base for network
    simulation and teaching environments: for example, see
    \cite{Pizzonia:2008:netkit,krap:2004,davoli:2005:vde}
  } without requiring special running privileges.
\item \ac{UML} enforces limits on the resource usage: e.g., a \ac{UML}
  system will use no more than the memory that has been assigned to
  it.
\end{itemize}

\ac{UML} ensures process and kernel address space separation in the
guest system through a mechanism called ``SKAS0'' (see Section
``\protect\ac{UML} Execution Modes'' on page~128 of \cite{uml} for
details).

\section{Architecture and usage}
\label{sec:architecture}

An AppPot appliance appears to a user as consisting of a few files:
\begin{enumerate*}[\em (1)]
\item an \emph{AppPot disk image}, which is a complete GNU/Linux system
  installed in a partition image file (in ``raw'' format);
\item an \acs{UML} Linux kernel;
\item a shell script \texttt{apppot-start} used to run a command-line
  program within the AppPot appliance;
\item a few auxiliary programs that enable optional features of
  AppPot (networking, I/O streams redirection).
\end{enumerate*}
All these files can, all or in part, be installed system-wide so that many
users can benefit from a shared installation.

Most of the components of AppPot only interact during the Linux boot
process.  The sequence of steps taken by an AppPot appliance from
invocation to the execution of a user-specified command is the
following.

{\em 1.} Users invoke the {\texttt{apppot-start}} script, optionally
  specifying a command to run in the AppPot appliance.
  Command-line options allow to set the path to the raw disk image
  file and the \ac{UML} kernel, or specify \ac{UML} boot arguments
  like maximum virtual memory.

{\em 2.} The \ac{UML} kernel ---running as a user process--- performs the
  normal Linux boot sequence, mounts the raw disk image file, and
  executes the startup scripts.  The Linux console I/O is redirected
  to the \ac{UML} process standard input and output streams.

{\em 3.} The {\texttt{apppot-init}} script is the last program executed
  as part of the boot sequence.  In detail, it does the following:
  \begin{enumerate}[\em a.]
  \item It reads the kernel boot command-line, and recognizes
    specially-formatted arguments put there by \texttt{apppot-start}.
    (For instance, the path to the ``changes'' archive file.)
  \item\label{init:mount} Mounts the current working directory of
    \texttt{apppot-start} in the \ac{VM} filesystem, and uses it as a
    working directory for all subsequent steps.
  \item\label{init:uid} Alters the UNIX \acs{UID} and \acs{GID} of the
    regular user in the AppPot appliance to match those of the owner of
    the working directory in the host filesystem.\footnote{%
      This is necessary as accesses to the host filesystem undergo
      ordinary access control by the host kernel; operations performed
      by processes running in the \ac{UML} \acl{VM} appear as done by
      the user running the \ac{UML} process in the host system.  }
  \item\label{init:changes} If a ``changes'' archive file is
    specified, merges its contents into the currently-running \ac{VM}
    filesystem. (See Section~\ref{sec:usage-development} for an more
    details and an example.)
  \item\label{init:exe} Commands to run non-interactively in the
    AppPot appliance can be given on the \texttt{apppot-start}
    command-line, or specified in a script named ``apppot-run'' which
    has to be placed in the host directory where AppPot is run.  (For
    example, the ``apppot-run'' script can pre-process the input data
    and then execute the main application.)

    So, if a command is specified on the boot command line,
    \texttt{apppot-init} runs it; otherwise, it checks if a startup
    script exists in the appliance or in the working directory. 

    If none of the above applies, \texttt{apppot-init} starts an
    interactive shell on the system console.
  \item When the command in the previous step has terminated,
    \texttt{apppot-init} initiates the shutdown sequence.
  \end{enumerate}

Communication between the \texttt{apppot-start} script (invoked by the
appliance users) and the \texttt{apppot-init} one (which drives the
\ac{VM} boot process) happens through the kernel boot parameters.
This is a very generic mechanism, that can easily be extended to drive
AppPot appliances with any virtualization \ac{API} that supports
some kind of argument passing between the caller and the \ac{VM}.  For
instance, the ``user-data'' mechanism of the \acs{EC2} \acs{API} could
serve this purpose as well. 

The reference AppPot image is based on the stable release of the
Debian distribution.  Apart from being optimized for running a
single-user task in a virtualized environment (i.e., multi-user
support, hardware-related software and most daemons have been turned
off, resulting in a total startup time of 9s), it is a regular Debian
install.

Note that, since the AppPot system image is a raw disk image file, it
can be run through any virtualization software that can read this disk
format (e.g., \acs{KVM} or Xen); so it can also be used on
infrastructures that support full virtualization (i.e. \acs{IaaS}
cloud), or on non-Linux hosts.
On the other hand, running an AppPot appliance as a batch job through
the \texttt{apppot-start} script is currently supported with \ac{UML}
only. 
While support for any virtualization system accessible with the
\emph{libvirt} \acs{API} \cite{libvirt:website} could easily be added,
starting a virtual machine is a privileged operation with most
virtualization systems, and it would not be safe to allow any user to
do that, since resource allocation is not handled by the hypervisor
software.\footnote{%
  Indeed, in many consolidation use cases, a physical machine's
  resources are oversubscribed by starting more \acp{VM} then the
  underlying hardware could actually run at the same time.
}

Data movement in- and out of the \ac{VM} happens through the shared
filesystem: in particular, this allows AppPot to run any command in a
working directory that is shared with the host system, so that
invocation of a command through \texttt{apppot-start} is completely
transparent to the user.  However, this mechanism is
\ac{UML}-specific; if a different virtualization system is to be used,
alternatives should be implemented to move data in and out of the
AppPot appliance.


\subsection{Using AppPot}
\label{sec:using}

Users receive an AppPot system image, containing a working
installation of a GNU/Linux distribution.  
Users have full access rights to the AppPot system image thus they can modify
it by installing new software, libraries, reference data e.g.,
their own version of a computational code or a reference dataset that
will be used during the computational analysys.  The choice of Debian as
the base distribution plays a role here, as there are already several
thousand packages available in the Debian main archive, including
several popular scientific codes \cite{debian-science:website}. 

There are three main usage modes of AppPot, detailed below.

\paragraph{Interactive local execution} In this case AppPot is
started on a local machine; the disk image file as well as input data are
directly available. Typical use cases of this usage mode are code validation
and appliance customization: users start the AppPot appliance invoking
``apppot-start'' from the shell passing the path to the disk image
file, \ac{UML} kernel and, if necessary, local filesystem location of input
data. At the end of the boot process, an interactive shell is opened
on the system console.

\paragraph{Batch job on a cluster resource} This is a more common
case for data analysis and scientific code execution. In a typical
cluster setup, appliance image file, \ac{UML} kernel and input data
are made available to the batch cluster execution node; the
\texttt{apppot-start} command is invoked by the batch job to run a
command non-interactively within the AppPot appliance.

AppPot execution is monitored through the batch job stdout: the AppPot
startup script takes care of redirecting the appliance console output
onto the standard output of the batch job.

\paragraph{Grid job} AppPot can also be executed as a grid job on a
distributed infrastructure. In this case, the disk image file,
execution script and reference data need to be transferred to the
destination node before the execution. This is normally achieved by
specifying those input files as part of the grid job description
file. Similarly to the batch job execution case, appliance execution
on the grid can be monitored throught the grid job's standard output
stream.

If the base AppPot image is already deployed at the execution site,
network traffic can be reduced by sending just a changes archive file
with the differences between the base AppPot and the user-modified one.

\section{Real-world usage}
\label{sec:usage}

This section illustrates how AppPot can be used to address the issues
presented in Section~\ref{sec:usecases}. As a production environment,
AppPot instances have been run as grid jobs on the distributed and
heterogeneous Swiss National Grid Infrastructure \acs{SMSCG}.

\subsection{Deployment of complex applications}
\label{sec:usage-deployment}

In Section~\ref{sec:usecase-deployment}, we identified three
issues that can make application deployment on a cluster a complex
sysadmin task.  

A common solution to issue \ref{R3} is to deploy only appliances that
have undergone a ``certification'' process by the local security
personnel or other trusted entity.  AppPot solves the issue by
allowing the execution of the appliance through \ac{UML}; there is
thus no need to supervise the appliances installed on the system, as
they can gain no more privileges than the executing user already can.
Actually, systems administrators can just provide the minimal \ac{UML}
infrastructure (the kernel binary, the \texttt{slirp} executable,
etc.) and allow users to install their own appliances.

Issues \ref{R1} and \ref{R2} can be mitigated by virtualized
appliances of any kind; the specific contribution of AppPot in this
case is to enable the use of such appliances on \emph{existing}
batch-oriented computing infrastructures.

In addition, AppPot allows the use of software appliance in a
non-interactive fashion, and especially in traditional UNIX
command-line scripting environments.  In particular, this enables the
integration of AppPot appliances in data analysis pipelines and other
automated data processing systems.

As an example, we consider the use case illustrated in
\cite{costantini+murri:iccsa2012}: ABC is a computational chemistry
code, that takes as input a \ac{PES} specification, in the form of a
Fortran function.  For computational efficiency, this \ac{PES}
function is compiled together with the rest of the ABC source code to
form an executable binary specific for a certain molecule.  However,
the ABC build procedure requires the G95 Fortran compiler
\cite{g95:website}, which is not part of any common GNU/Linux
distribution.  AppPot allows to create an ABC appliance by extending
the base image with the needed G95 compiler together with the ABC
source code, and an ``auto run'' script that looks for the \ac{PES}
file, compiles and bundles it into the ABC binary, and then run the
resulting executable with the user-supplied parameters.  The resulting
appliance provides a solution for ABC, that preserves the flexibility
of the original code, but has no dependencies and thus can be deployed
and run on any GNU/Linux cluster.  In particular, this has been used
in the cited \cite{costantini+murri:iccsa2012} to implement a
grid-based analysis workflow.

As a further example, consider the generation of plots and graphs to
create synthetic views of the analyzed data. Traditionally, the
generation of plots and reports has been done ``offline'' on desktop
machines, or on separate visualization clusters. However, with the
increase of the volume of processed data, this is no longer feasible,
and there is a growing request to use the computing facilities to
generate such plots and reports.  But no single graphing library or
system has yet emerged as a widely-used standard, which implies that a
good fraction of users will not find their favorite graphing library
pre-installed on the computing infrastructure they have access to.
Using AppPot, a user can install its favourite plotting library and
its dependencies (e.g., Python~2.7 with
Matplotlib~\cite{matplotlib:website}) and run the post-processing
plotting step as a regular computational job.

\subsection{Running development code}
\label{sec:usage-development}

The gist of requirements \ref{R4}--\ref{R7} is that users should be
able to quickly create new AppPot appliances and deploy those on the
execution sites without the need for sysadmin support. Here we show
how this can be accommodated by the tools provided by AppPot.

Recall from Section~\ref{sec:using} that the AppPot disk image is a
regular file in the host filesystem, and that AppPot users have full
control on it.  Indeed, they can arbitrarily modify a running disk
image using standard GNU/Linux administration commands in the
appliance.  It is thus easy to make several copies and modify each of
them independently.  This clearly satisfies requirements~\ref{R6}
and~\ref{R7}.

Requirement \ref{R5} can be addressed using the snapshot/changes
implemented in AppPot: users can create a ``changes'' file that
encodes the differences of the locally-modified appliance with a
``base'' one. During AppPot boot, the \texttt{apppot-init} script will
re-create the modified appliance from the base one, by merging in the
changes.

The ``changes'' files are produced with the \texttt{apppot-snap}
utility.  It should be invoked a first time in order to mark a
``base'' system: it records the state of all files in the appliance.
This ``base'' system is then deployed in a location on the execution
cluster, shared by all users.

When a user has finished modifying its local copy of the AppPot
appliance, invokes \texttt{apppot-snap} again.  Each time it is
invoked with the \texttt{changes} subcommand, \texttt{apppot-snap}
compares the state and content of each file with the recorded state,
and creates an archive file with all changed content. 

This archive file can then be merged into a different appliance that
has the same ``base'' content.  Therefore, users can just send the
``changes'' file along with their grid jobs, to re-create their local
AppPot environment on the remotely-installed one.  This clearly
minimizes the time to deployment of a modified appliance.

\subsection{Dynamical expansion of clusters}
\label{sec:cloudbursting}

An AppPot-based compute node \ac{VM} has been used in the \acs{VM-MAD}
to provide dynamic expansion of a computational cluster using
virtualized computing hardware.  In \acs{VM-MAD}, an AppPot instance
is submitted to the \acs{SMSCG} infrastructure each time a site seeks
to expand its resources; the AppPot instance has been customized by
the site admin to be a replica of the standard compute node, that
connects back to the home site using a \acs{VPN}.

This practice is generally known as ``cloudbursting''
\cite{defn-cloudbursting}, when the additional resources are drawn
from a public Cloud (e.g., Amazon's \acs{EC2}).  

The use of AppPot appliances as batch cluster compute nodes, allows an
institution computational cluster to be expanded using nodes from
other resources in the same infrastructure, without its users
realizing that the are using grid computing at all.  This is similar
to the ``glide-in'' mechanism implemented in the Condor batch
execution system \cite{thain+livny:2005:}.

\section{Limitations}
\label{sec:limitations}

Usage of AppPot has also shown some unexpected
limitations and issues, mostly with the attempt of doing a completely
user-controlled and user-initiated deployment of \acp{VM}.

Most issues in the actual usage of AppPot appliances originate from
the way \acs{UML}'s ``SKAS0'' mode operates: \ac{UML} must ensure
address space separation of processes running in the virtualized
system, but this is normally achieved by an \ac{OS} by using
hardware-assisted protection of memory pages, which is a privileged
operation.  So the \ac{UML} author found this solution:
\begin{enumerate}[\em a.,ref={\em \alph*.}]
\item \label{item:1}
  Each execution context\footnote{%
    In the Linux kernel, processes and threads are both instances of
    a more general ``execution context'' concept.
  } in the guest system is actually a separate process in the host
  system.
\item \label{item:2}
  For each memory page needed by the guest kernel, a page of memory
  is allocated from the host kernel.
\item \label{item:3}
  All memory pages that are shared among processes in the guest system
  are directly mapped (via a
  \href{http://fscked.org/writings/SHM/shm-2.html}{\texttt{mmap()}}
  call) to a segment of data in a temporary file.
\end{enumerate}
This effectively offloads address space separation to the host kernel
(by~\ref{item:1}) while still retaining the possibility of sharing
portions of memory (by~\ref{item:2} and~\ref{item:3}), which is
necessary because all processes in the virtual system must see the
same kernel image.

Therefore, a \ac{UML} process makes a large number of \texttt{mmap()}
calls to the host system (one for each page of memory handed out by
the guest kernel).  As it turns out, the default limit on the number
of \texttt{mmap}-ed segments of memory, bounds the maximum memory that
a \ac{UML} kernel can allocate to 256MiB, which is too low for any
computational usage.

In addition, all these memory pages are backed by temporary ``shared
memory'' file storage; therefore, the available space on the
filesystem where these files are created limits the total amount of
memory available to AppPot/\ac{UML} \acp{VM}.  

In both cases, it is a straightforward systems administration task to
change the default so that \ac{UML} \aclp{VM} can allocate a larger
portion of memory.  As simple as they are, especially compared to
installation and support of arbitrary libraries and applications,
these configuration settings still represent an enabling step that
systems administrators must perform in order to allow AppPot
appliances on a computational cluster: users cannot do everything by
themselves.\footnote{%
  For the sake of exactness, only altering of the kernel settings is a
  privileged operation: changing the backing filesystem for \ac{UML}
  memory can be done by any user via an environment variable.
  However, this implies that users know where a suitable filesystem is
  located on each cluster, which assumes a more detailed knowledge
  about the cluster setup than would be desirable, not to mention that
  in a large grid infrastructure to gather this information for each
  and every cluster is a nontrivial task in itself.  }

A different kind of issue originates in the way batch systems enforce
job memory limits. This issue has been observed with Sun/Oracle Grid
Engine 6.2 and \acs{PBS}/\acs{TORQUE}; it's very likely to happen with
other batch systems as well.

Recall that each process and thread in the \ac{UML} \ac{VM} is
actually running as a process in the host system. These processes share
part of their memory space (the whole kernel and its data structure);
in the case of threads they actually share most of it.  However, it
seems that batch systems assume that much of the memory used by a
process is private and only a negligible fraction is shared; therefore
they wrongly reckon the total memory usage by \ac{UML} processes by
summing the memory occupation without accounting for shared pages.
Hence, almost \ac{UML}-based appliances quickly hit the memory limit
and are killed by the batch system.

Circumventing this bug is technically simple (turn off enforcement of
memory limits for \ac{UML} batch jobs), but this can be a significant
policy change in the management of the cluster, which brings back the
negotiations with the centralized cluster administration that \ac{UML}
virtualization was meant to avoid.

While \ac{MPI} communication among AppPot instances is possible
(although this is matter for future development), \ac{UML} lacks
support for \ac{SMP}, which limits its use in multi-threaded
applications.

\section{Related work}
\label{sec:related}


A related attempt to use virtualization in the world of scientific
computing as been made by \acs{CERN} in order to ``allow end-users to
effectively use their desktops and laptops as for analysis and Grid
interface'' \cite{buncic:2008:cernvm}.  The CernVM virtual machine
image can run under the Xen or \acs{KVM} hypervisors and contains a
minimal operating system installation to run the \acs{LHC} experiment
software as well as to function as an interface host for the
\acs{WLCG} grid infrastructure; access to the actual experiment
software and grid middleware happens through an \acs{HTTP}-based
filesystem: the real application files are stored on a server farm at
\acs{CERN}. There are thus two major differences to AppPot:
\begin{itemize}\item 
  The CernVM is itself a specific software appliance, not intended for
  customization or redistribution by users.
\item 
  The way CernVM bridges the local user environment and the
  batch-execution one ``goes in the reverse direction'' relative to
  AppPot: CernVM lets users run programs in the same environment they
  find in the grid systems, whereas a use case for AppPot is to let
  the batch jobs environment be customized and prepared by the user.
\end{itemize}

The use of virtualization in cluster environments has been considered
in \cite{emeneker-stanzione:hpc-uml}; however, the authors approached
the usage of the Xen and \ac{UML} virtualization systems in order to
provide ``virtual clusters'' out of a local pool of general-purpose
hardware.  A somewhat similar use case for AppPot is discussed in
Section~\ref{sec:cloudbursting}.

The use of \ac{UML} for \ac{HPC} and distributed applications is also
the subject of \cite{jiang-dongyan:violin,ruth-jiang-xu-goasguen}; the
aim of the authors is rather to provide ``a middleware system
\emph{[...]}  to support mutually isolated virtual distributed
environments in shared infrastructures'', and the focus is on
supporting applications that do not fit the batch computing or the
service model.

The ``application virtualization'' technique seems to be more
well-established in the Windows operating system community (see:
\cite{wikipedia:application-virtualization,pietroforte:application-virtualization})
than it is in the Linux one.  However, the authors of this paper know
of no earlier attempt to bring the ``software appliances'' concept
into batch-oriented computing.

\section{Conclusions}
\label{sec:conclusions}

This paper has presented a system to bring the benefits of
virtualization-based \acs{IaaS} clouds to existing batch-oriented
computing infrastructures. 

AppPot is currently used in production within the Swiss National Grid
Infrastructure \acs{SMSCG}, supporting several use cases like those
presented in this paper.  We are collecting feedback on the
effectiveness of AppPot in large-scale grid computations; we would
like to stress that such effectiveness is not just a function of
system performance, but should also include consideration of how it
makes large-scale computing more accessible (on the users' side) and
manageable (on the systems administrators side).


Some technical improvements could spark more widespread adoption; they
are discussed in the following subsection.

\subsection{Future development}
\label{sec:future}

Building on another feature of \ac{UML}, namely, tunneling all \ac{IP}
traffic through a local \ac{UDP} multicast socket, it would be
possible to use AppPot appliances to run \acs{MPI} jobs.  The
\texttt{apppot-start} script would recognize a parallel environment,
and launch an AppPot instance for every \acs{MPI} rank, with the
appropriate network parameters so that the instances can communicate
over the multicast channel.  The combined action of virtualization and
tunneling could possibly be a heavy performance hit on the \acs{MPI}
subsystem, but the increase in flexibility and the other advantages of
virtualized appliances could still make this an attractive solution in
some cases.


Finally, as explained in Section~\ref{sec:architecture}, the generic
architecture of AppPot can easily be adapted to run on a variety of
virtualization systems.  We believe that such an extended AppPot can
be an effective construction ingredient in mixed Grid/Cloud
infrastructures.

\bibliographystyle{abbrv}
\bibliography{grid.bib}

\appendix

\section{List of acronyms}
\label{sec:acro}
\input{acronyms}

\end{document}

%% file: acronyms.tex


\acused{API}%
\acused{HTTP}%
\acused{SWITCH}%
\acused{URL}%

\begin{acronym}[MMMMMMM]
  \acro{AAI}{Authentication and Authorization Infrastructure}%
  \acro{AC}{Attribute Certificate\acroextra{ (VOMS, X.509)}}%
  \acro{API}{Application Programming Interface}%
  \acro{ARC}{Advanced Resource Connector}
  \acro{CA}{Certification Authority}%
  \acro{CERN}{Centre Européen pour la Recherche Nucléaire\acroextra{
      (European Center for Nuclear Research)}}%
  \acro{COMPCHEM}{Computational Chemistry}
  \acro{CPU}{Central Processing Unit}
  \acro{CSR}{Certificate Signing Request}%
  \acro{DAG}{Directed Acyclic Graph}
  \acro{DN}{Distinguished Name}%
  \acro{EC2}{Amazon's Elastic Compute Cloud}%
  \acro{ECP}{Enhanced Client or Proxy}%
  \acro{EGEE}{Enabling Grids for E-sciencE}%
  \acro{EGI}{European Grid Initiative}
  \acro{FQAN}{Fully-Qualified Attribute Name\acroextra{ (VOMS)}}%
  \acro{GAMESS}{General Atomic and Molecular Electronic Structure System\acroextra{ (a Computational Chemistry application, see \cite{GAMESS:1993,GAMESS:2005})}}%
  \acro{GASuC}{Grid Applications Porting Group}
  \acro{GC3}{Grid Computing Competence Center, University of Zurich}
  \acro{GEMS}{Grid-Enabled Molecular Simulator}
  \acro{GID}{Group Identifier}
  \acro{HPC}{High-Performance Computing}
  \acro{HTC}{High-Thorughput Computing}
  \acro{HTTP}{HyperText Transfer Protocol}%
  \acro{KVM}{Kernel-based Virtual Machine}%
  \acro{IaaS}{Infrastructure as a Service}
  \acro{ID-WSF}{Identity Domain - Web Service Framework\acroextra{ (Shibboleth)}}
  \acro{IGI}{Italian Grid Initiative}
  \acro{IGTF}{International Grid Trust Federation}%
  \acro{IdP}{Identity Provider\acroextra{ (Shibboleth)}}%
  \acro{IP}{Internet Protocol}
  \acro{LHC}{Large Hadron Collider}%
  \acro{MPI}{Message Passing Interface}%
  \acro{NGI}{National Grid Infrastructure}
  \acro{OS}{Operating System}\acused{OS}%
  \acro{PaaS}{Platform as a Service}
  \acro{PBS}{Portable Batch System}
  \acro{PES}{Potential Energy Surface}
  \acro{PKI}{Private Key Infrastructure}%
  \acro{POSIX}{Portable Operating System Interface}%
  \acro{SaaS}{Software as a Service}
  \acro{SAML2}{Security Assertion Markup Language version 2}%
  \acro{SAML}{Security Assertion Markup Language}%
  \acro{SLCS}{Short-Lived Credential Service}%
  \acro{SMP}{Symmetric Multi-Processing}%
  \acro{SMSCG}{Swiss Multi-Science Computational Grid}
  \acro{SOA}{Service-Oriented Architecture}
  \acro{SP}{Service Provider\acroextra{ (Shibboleth)}}%
  \acro{SSO}{Single Sign-On}%
  \acro{SWITCH}{Swiss Academic Network Provider}%
  \acro{SWITCHaai}{Swiss Federated AAI}\acused{SWITCHaai}%
  \acro{TORQUE}{Terascale Open-source Resource and QUEue manager}
  \acro{TCP}{Transmission Control Protocol}\acused{TCP}%
  \acro{TERENA}{Trans-European Research and Education Networking
    Association}%
  \acro{UDP}{User Datagram Protocol}\acused{UDP}%
  \acro{UI}{User Interface}
  \acro{UID}{User Identifier}
  \acro{UML}{User-Mode Linux}
  \acro{URL}{Uniform Resource Locator}%
  \acro{VHO}{Virtual Home Organisation}%
  \acro{VM}{Virtual Machine}
  \acro{VM-MAD}{Virtual Machines Management and Advanced
    Deployment\acroextra{ (project ETHZ.7 funded by the SWITCH AAA track)}}%
  \acro{VOMS}{Virtual Organisation Membership Service}%
  \acro{VO}{Virtual Organization}
  \acro{VPN}{Virtual Private Network}
  \acro{WLCG}{Worldwide {LHC} Computing Grid}%
  \acro{WMS}{Workload Management System\acroextra{ (one
      component of the gLite grid middleware system)}}
  \acro{XACML}{eXtensible Access Control Markup Language}%
  \acro{xRSL}{Extended Resource Specification Language}
\end{acronym}
